\documentclass[12pt]{article}
\usepackage{epsfig}

\newcommand\pubnumber{CERN-TH/2001-009 \\ NYU-TH/01/01/01}
\newcommand\pubdate{\today}

\def\nostro{ {(a) Physics Department, New York University ,}\\
  {4 Washington Place, New York, NY, 10003, USA.}
  \\\vspace{.3cm}
  {(b) Theory Division, CERN,} \\
  {CH-1211, Geneva 23, Switzerland.}}

\def\Title#1{\begin{center} {\Large #1 } \end{center}}
\def\Author#1{\begin{center}{ \sc #1} \end{center}}
\def\Address#1{\begin{center}{ \it #1} \end{center}}

\newcommand\pubblock{\rightline{\begin{tabular}{l} \pubnumber\\
         \pubdate  \end{tabular}}}
\newenvironment{Abstract}{\begin{quotation}  }{\end{quotation}}

\def\Acknowledgements{\bigskip  \bigskip \begin{center} \begin{large}
             \bf ACKNOWLEDGEMENTS \end{large}\end{center}}
\def\g{\Gamma} 
\def\gg{{\rm I}\!\g}

\newcommand{\gh}{\hat{\g}}

\def\beq{\begin{equation}}
\def\eeq#1{\label{#1}\end{equation}}
\def\eeqn{\end{equation}}

\def\beqa{\begin{eqnarray}}
\def\eeqa#1{\label{#1}\end{eqnarray}}
\def\eeqan{\end{eqnarray}}

\let\bar=\overbar

\def\Dslash{\not{\hbox{\kern-4pt $D$}}}
\def\dslash{\not{\hbox{\kern-2pt $\del$}}}

\def\msb{{\bar{\ssstyle M \kern -1pt S}}}

\begin{document}
\begin{titlepage}
\pubblock

\vfill
\Title{\bf 
 On the two-loop electroweak amplitude\\
  of the muon decay\footnote{To appear in the Proceedings of the 
5th International Symposium on Radiative Corrections (RADCOR-2000) 
Carmel CA, USA, 11--15 September, 2000. Talk by P.~A.~G..
}
}
\vfill
\Author{P.A. Grassi$^a$ and T. Hurth$^b$
}
\Address{\nostro}
\vfill
\begin{Abstract}
     We present an analysis of the two-loop amplitude of 
      the muon decay in the Standard Model (SM) using algebraic renormalization techniques. 
      In addition, we discuss  a manifestly BRST invariant IR regulator 
      for the photon within the SM.
\end{Abstract}
\vfill
%\begin{Presented}
%5th International Symposium on Radiative Corrections (RADCOR-2000) \\
%Carmel CA, USA, 11--15 September, 2000
%\end{Presented}
%\vfill
\end{titlepage}
\def\thefootnote{\fnsymbol{footnote}}
\setcounter{footnote}{0}

\section{Introduction}
\label{sec:introduction}
In perturbative multi-loop calculations, the subtraction of 
UV-divergences in quantum field theory generally leads to Green functions 
which fail to respect the symmetries of the theory. With the exception of the well-known 
$\gamma_5$ problem, the method of dimensional regularization is compatible
with the gauge  symmetry but it breaks the supersymmetry. 
In these cases a practical method is needed to restore the symmetry
identities of the gauge symmetry or of the supersymmetry. 
Here the method of algebraic renormalization \cite{pige} 
supplies a complete solution. 
However, this method has rarely been used in practical
calculations although it has been applied intensively in order to
demonstrate the renormalizability of various models.

In a recent papers~\cite{amt_1,amt_3}, we reviewed the method of 
algebraic renormalization from a practical point of view and proposed 
an algebraic method combining the advantages of the background field method 
(BFM) and the simplification of (intermediate) Taylor subtractions. 
The method is independent of the regularization scheme;
since the local breaking terms are under control, one can use the 
most convenient regularization scheme in a specific application.             
After a straightforward analysis of the corresponding (Ward-Takahashi Identities) WTIs and 
(Slavnov-Taylor Identities) STIs,
the spurious anomalies introduced by a non-invariant regularization 
scheme were  shown to reduce to a few universal breaking terms which 
depend only on finite Green's functions. 
The method was already  applied to several phenomenologically 
relevant examples in the SM, such as the two-loop contributions to the 
processes $H \rightarrow \gamma \gamma$, to $B \rightarrow X_s \gamma$
and to the three-gauge boson vertices~\cite{amt_1,amt_2,amt_3}. 

Because of the experimental precision of standard model observables
at  LEP (CERN, Geneva), at SLC (SLAC, Stanford) and at  TEVATRON
(FERMILAB), calculations of quantum corrections 
on the two-loop level are necessary; and the 
$\gamma_5$ play a critical role here.
The purpose of this note is to offer a theoretical analysis of 
the electroweak two-loop contribution to the muon decay
using our algebraic method, and to show its efficiency.

Since a detailed self-contained discussion of the fundamental symmetry 
constraints for the SM, of the algebraic renormalization
procedure in the BFM, and of our specific subtraction  
method can be found in~\cite{amt_1,amt_3}, 
we restrict ourselves here to reviewing briefly  the basic steps 
of our method (Sec. 2).
Then we discuss the muon-decay amplitude, in particular the 
gauge-invariant subset of two-loop diagrams which is sensitive 
to the $\gamma_5$ problem (Sec. 3).
It is well-known that there is a physical infrared divergence 
present in the muon-decay amplitude.
For this purpose, we propose an IR regulator that is manifestly compatible 
with all the symmetries of the SM (Sec. \ref{IR}).

%%%%%%%%%%%%%%%%%%%%%%%%%%%%%%%%%%%%%%%%%%%%%%%%%%%%%%%%%%%%

\section{General Strategy}

In the following, we briefly review the main steps elaborated
in~\cite{amt_1, amt_3} to renormalize a gauge model with a non-invariant 
regularization technique. The BFM turns out to be 
very important for our purposes and, therefore, we 
quantize the SM in the `t Hooft background gauge~\cite{bkg,msbkg,grassi}. 

The use of a non-invariant regularization scheme 
induces breaking terms into the STIs
\begin{equation}
  \label{STI}
  {\cal S}\left( \g^{(n)} \right)  = \hbar^n \Delta^{(n),S} + 
  {\cal O}(\hbar^{n+1})\,, 
\end{equation}
which implement the Becchi-Rouet-Stora-Tyutin (BRST)
symmetry, and into the WTIs
\begin{equation}
  \label{WTI}
  {\cal W}_{(\lambda)}\left( \g^{(n)} \right)  = \hbar^n \Delta^{(n),W} +
  {\cal O}(\hbar^{n+1})\,, 
\end{equation} 
which implement the background gauge invariance of the SM. 
The definition of ${\cal S}$ and ${\cal W_{(\lambda)}}$ 
is given in~\cite{amt_1}. 

The local breaking terms are denoted by $\Delta^{(n),S}$ and $\Delta^{(n),W}$.
Note that the locality is a consequence of the Quantum Action Principle. 
Here and in the following, $\Gamma^{(n)}$ denotes the $n$-loop order
regularized and (minimally) subtracted one-particle-irreducible (1PI) function.
$\Gamma^{(n)}$ includes the renormalization of all subdivergences. 
The STIs and the WTIs are not able to fix the Green functions completely.  
Indeed it is possible to add invariant local terms to the action changing the
normalization conditions of the Green functions. A complete analysis on the  
normalization conditions for the SM can, for instance, be found
in~\cite{grassi}.

Acting on the broken WTIs (\ref{WTI}) 
with the Taylor operator $(1-T^\delta)$ one gets
\begin{equation}
  \label{WTI.1}
  (1-T^\delta)  {\cal W}_{(\lambda)} \left( \g^{(n)} \right) 
  \,\,=\,\,0  \,,
\end{equation} 
where 
$\delta$ has to be chosen in such a way that 
$(1-T^\delta)\Delta^{(n),S/W}=0$.
After commuting the Taylor operator 
$(1-T^\delta)$ with ${\cal W}_{(\lambda)}$,  
we obtain
\begin{equation}
  \label{WTI.2}
  {\cal W}_{(\lambda)} \left[(1-T^{\delta'} ) \g^{(n)} \right]  
  \,\,=\,\,
  \left[T^{\delta}{\cal W}_{(\lambda)} - {\cal W}_{(\lambda)}T^{\delta'} \right] \g^{(n)} 
  \,\,\equiv\,\, 
  \hbar^n \Psi^{(n),W}(\lambda) 
  \,,
\end{equation} 
where $\delta'$ is the naive power-counting degree
of $\g^{(n)}$. In general, one has $\delta  \geq   \delta'$,
hence the commutation of the Taylor operator with 
${\cal W}_{(\lambda)}$ leads to over-subtractions of 
$\g^{(n)}$ and, thus, to the new breaking terms $\Psi^{(n),W}(\lambda)$ 
occurring on the r.h.s. of Eq. (\ref{WTI.2}) (for more details see
\cite{amt_1,amt_3}). The breaking terms $\Psi^{(n),S}(\lambda)$ for the 
STIs are defined in the same way.  Therefore, the 
application of the Taylor subtraction 
on Eqs.~(\ref{STI}) and~(\ref{WTI}) transforms them into
\begin{eqnarray}\label{new_breal}
  {\cal S}\left( \gh^{(n)} \right) = \hbar^n \Psi^{(n),S} + 
  {\cal O}(\hbar^{n+1})\,
  &\mbox{and}&
  {\cal W}_{(\lambda)}\left( \gh^{(n)} \right)  = \hbar^n \Psi^{(n),S} 
  + {\cal O}(\hbar^{n+1})\,,
\end{eqnarray}
where $\hat\Gamma^{(n)} = (1-T^\delta)\Gamma^{(n)}$. 

The breaking terms $\Psi^{(n),S}$ and $\Psi^{(n),W}$ 
can be expressed in terms of a linear combination of ultra-violet (UV) finite  
Green functions.
Here we assumed that up to the $(n-1)$-loop order the Green functions are 
already correctly renormalized. 
The main difference between 
$\Psi^{(n),S}$ and $\Psi^{(n),W}$ is due to the linearity of the
corresponding operators ${\cal S}$ and ${\cal W}_{(\lambda)}$. 
In the former case one has to consider non-linear terms arising
from lower loop orders. On the contrary, the linearity of the WTIs 
enormously simplifies the evaluation of the breaking terms 
and of counterterms. 

Finally, we introduce
\begin{eqnarray}  
  \gg^{(n)}  &=& \gh^{(n)} - \Xi^{(n)} = (1-T^\delta) \g^{(n)} - \Xi^{(n)}\,,
  \label{eq:gg}
\end{eqnarray}
where $\Xi^{(n)}$ is chosen in such a way that  
the following identities are fulfilled:
\begin{eqnarray}
  {\cal S}\left( \gg^{(n)} \right) = 0\,,
  &&
  {\cal W}_{(\lambda)}\left( \gg^{(n)} \right) =0\,.
  \label{eq:conds}
\end{eqnarray}
In general, it is quite simple to compute the counter\-term, 
$\g^{C.T.} = T^\delta \g^{(n)} + \Xi^{(n)}$,
as it can be expressed in terms of Green functions expanded
around zero external momenta.

As already mentioned above, there is still the freedom 
to add invariant counter\-terms. In other words, we have the freedom to 
impose normalization conditions that lead in addition to
Eqs.~(\ref{eq:conds}) to the following equation being fulfilled:
\begin{eqnarray}
  {\cal N}_i \left( \gg^{(n)} \right) = 0\,,
  \label{nor_con}
\end{eqnarray}
where the index $i$ runs over all independent parameters of the SM.  
As the Green function $\Gamma^{(n)}$ also has to fulfill this condition
 we have for the counterterm
\begin{eqnarray}
  {\cal N}_i \left( T^\delta\g^{(n)} + \Xi^{(n)} \right)  &=& 0\,,
\end{eqnarray}
which is a local equation. This means that, whenever  
the effort to impose the normalization conditions is done 
the changes due to the  
subtraction are only a local changes which can be easily compensated.
For clarity let us consider an example: for the condition on 
the $W$ boson mass we could choose
${\cal N}_1(\g^{(n)}_{\hat W^+ \hat W^-}) = 
\g^{(n),T}_{\hat W^+ \hat W^-}(p^*) = 0$
where the superscript $T$ stands for the transverse part and 
${\rm Re}(p^*) = M_W$.
Notice that the imposition of normalization conditions is 
a very important ingredient for the computation in order to compare with  
other schemes and in order to simplify the breaking terms themselves. 

The procedure described so far is heavily based on the Taylor operator
$T^\delta$. In the presence of massless particles this may introduce infra-red
(IR) divergences. 
In~\cite{amt_2}  we presented a modified procedure
which resolves this spurious IR problem generally.  

\eject
\vfill
%%%%%%%%%%%%%%%%%%%%%%%%%%%%%%%%%%%%%%%%%%%%%%%%%%%%%%%%%%%%

\section{Muon Decay Amplitude}
\label{two-loop}

\subsection{General settings}

\begin{itemize}

\item {\it Muon decay amplitude} 

We want to focus on the $O(N_f \alpha^2)$ 
contributions including the two-loop three-point 
functions $\Gamma_{\hat{W}^+_\mu \bar{\nu} e}(p_\nu,p_e)$~\footnote{All momenta 
are considered as incoming. In the Green functions
$\Gamma_{\phi_1 \dots \phi_n}$ they are assigned to the 
corresponding fields starting from the right.
The momentum of the most left field 
is determined via momentum conservation. $\hat{W}$ 
denotes the background field corresponding to the 
quantum field $W$.} with a $\hat{W}$ and an electron (muon) and 
electron- (muon-) neutrino.
This subgroup of $O(N_f \alpha^2)$ contributions to the muon decay 
amplitude is the most delicate one regarding the $\gamma_5$-problem.
An example is given in Fig.~\ref{three-point}.

\begin{figure}
\begin{center}
\epsfig{figure=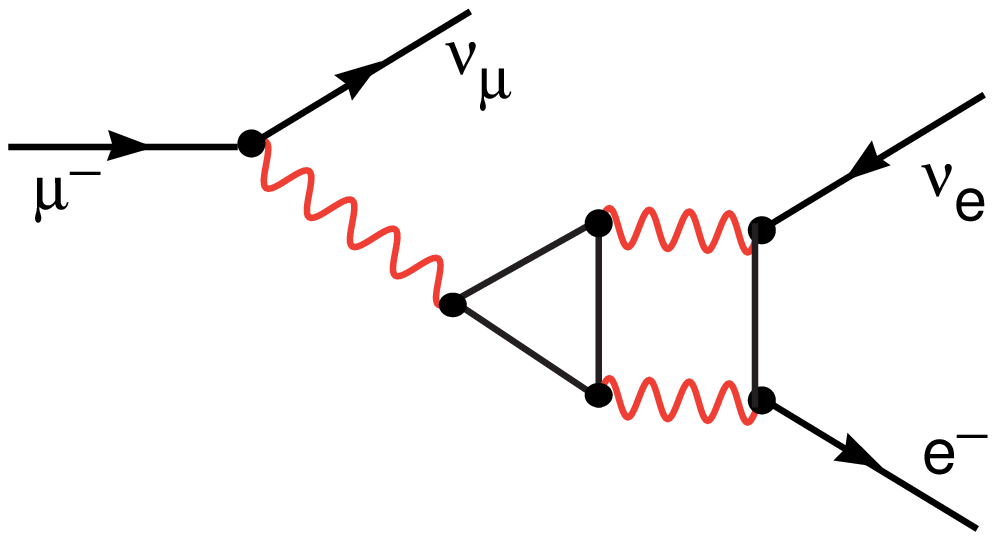,width=4.8cm}
\epsfig{figure=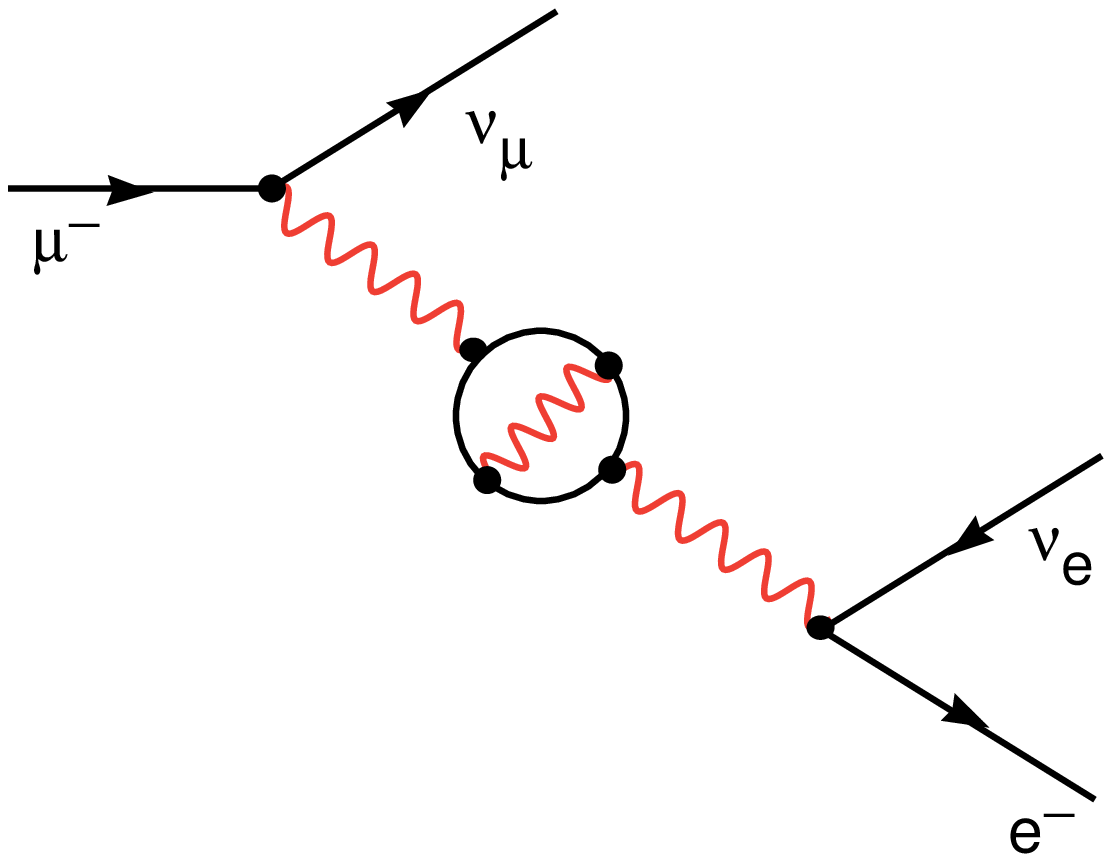,width=4.8cm}
\epsfig{figure=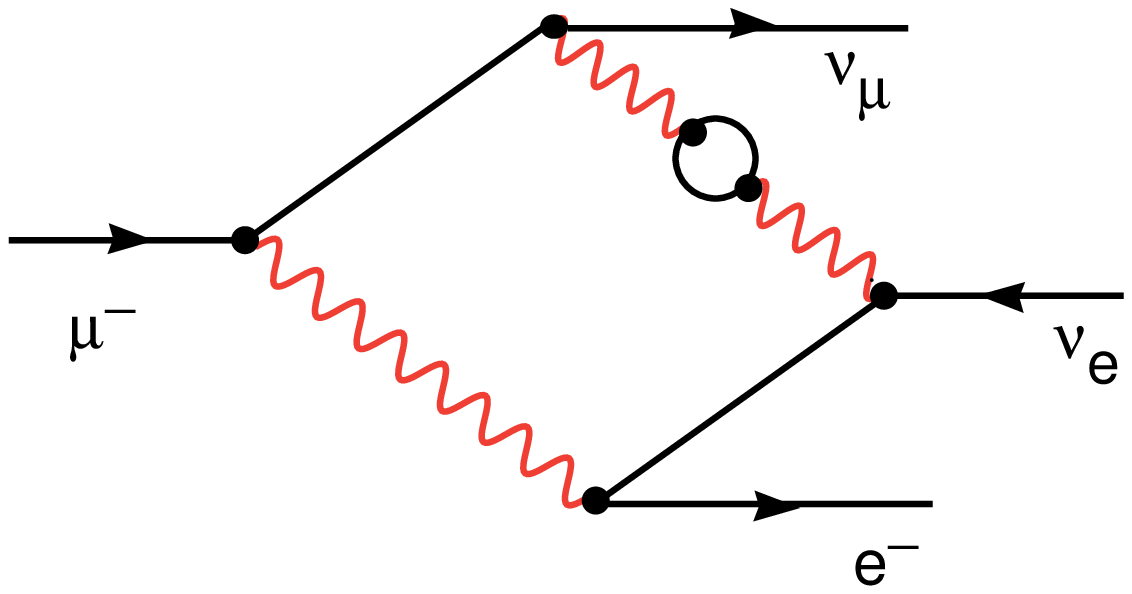,width=4.8cm}
\end{center}
\caption[f1]{
\small 
Example of $O(N_f \alpha^2)$ contributions to the muon decay 
amplitude with a two-loop three-point, two-loop two-point 
function and with box contribution} 
\label{three-point}
\end{figure}

There are further contributions at the two-loop level such as the 
diagrams including the two-loop-two-point function 
$\Gamma_{\hat{W}^+_\mu \hat W^-_\nu}(p)$ (Fig.~\ref{three-point}). 
These have already been discussed within our approach 
in \cite{amt_2}, but there are no problems  with $\gamma_5$ there. 
Moreover, there are two-loop box diagrams 
with a gauge-boson self-energy inside (as shown in the last diagram of 
Fig.~\ref{three-point}). 

Thus, let us focus on contributions like the one shown in
the first picture of Fig.~\ref{three-point}.
We have to consider the complete gauge-invariant subset 
of two-loop contributions to the three-point function
 $\Gamma_{\hat{W}^+_\mu \bar{\nu} e}(p_\nu,p_e)$.
Actually, there are various types of diagrams shown in 
Fig.~\ref{gaugesubset}.
Here we note that only the first two diagrams 
in Fig.~\ref{gaugesubset} change when switching from a 
conventional gauge to the `t Hooft-background gauge.

\begin{figure}
\hspace{0cm}
%\begin{center}
\epsfig{figure=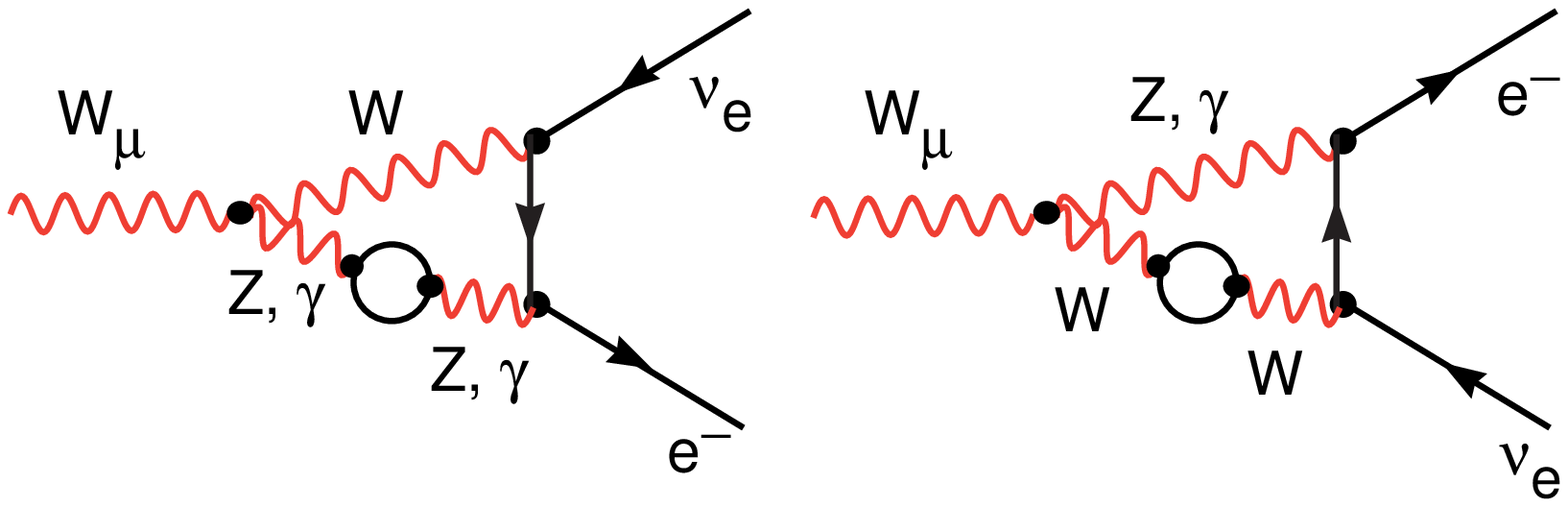,width=8cm} 

\vspace{-3.2cm}
\hspace{0cm}\epsfig{figure=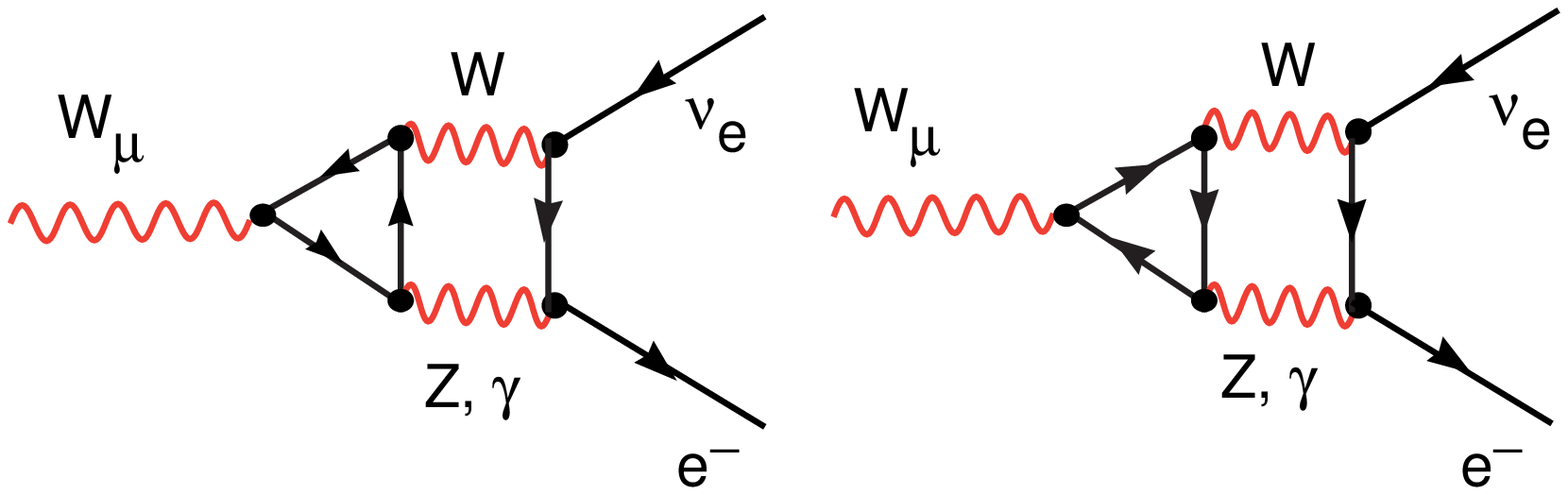,width=8cm} 
\hspace{0cm}\epsfig{figure=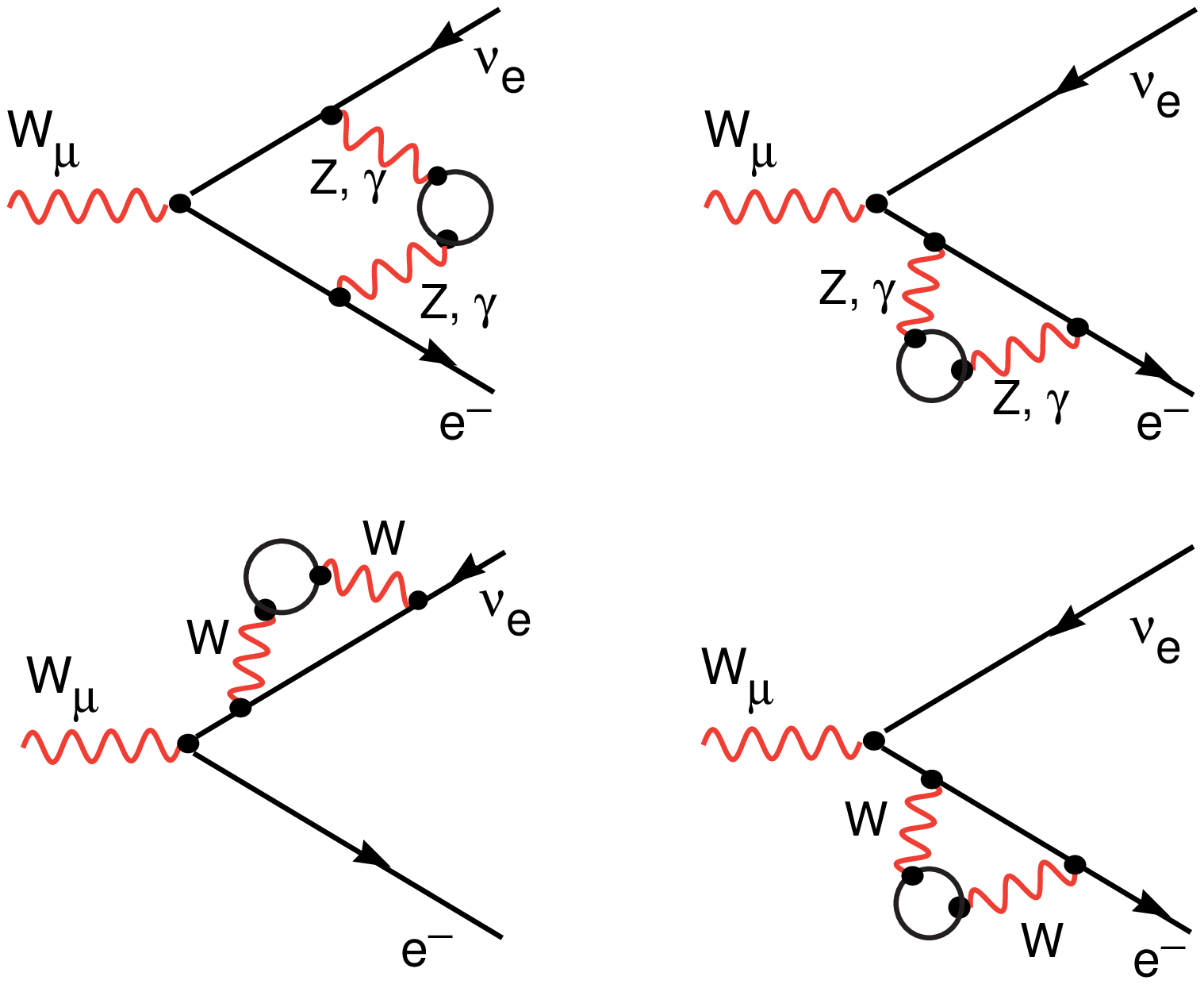,width=8cm} 
%\end{center}
\caption[f1]{\small Gauge-invariant subset of $O(N_f \alpha^2)$ contributions to 
the muon decay including vertex and external leg corrections.
}
\label{gaugesubset}
\end{figure}

\vspace{1cm}
\item{\it IR problems}

Among the $O(N_f \alpha^2)$ 
gauge-invariant subset of diagrams (see Fig.~(\ref{gaugesubset})), 
one has to consider those with a virtual photon. Those diagrams 
are potentially IR divergent and only the $O(N_f \alpha^2)$ contributions 
to the physical amplitude, after the inclusion 
of the Bremsstrahlung radiation of the 
external electron and muon, turn out to be finite. 

Historically \cite{sirlin}, the virtual photon contributions, 
namely the pure QED corrections, are analysed 
separately from the electroweak corrections 
by using a suitable decomposition of massless propagators. 
In that spirit, the pure QED corrections have been computed 
in \cite{stuart1,mat,ferroglia} and the remaining complete 
electroweak corrections are IR finite. In paper \cite{stuart2} 
the $O(N_f \alpha^2)$ corrections are computed in the MS 
scheme and the massless quark approximation has been used.  
In \cite{weig1,weig2} the exact fermionic contributions, in on-shell scheme, 
are taken into account. However, since 
we are not interested in the explicit evaluation of the 
muon amplitude, but only to present a procedure 
to handle the $\gamma_5$ problem in the present process, 
we will not disentangle the QED corrections in our 
considerations. 

As a consequence, we have to keep the possible IR divergences under 
control, namely we have to be sure that all the steps of the computation 
are IR regulated. The situation is worsened by the fact that, according 
to our procedure, the Taylor expansion in Eqs.~(\ref{WTI.2}) 
is performed at zero momentum. For those purposes, we propose a 
BRST invariant IR regularization for the photon within the SM (see 
Sec.~\ref{IR}).  This method regulates both physical and spurious IR divergences 
simultaneously. 

An alternative approach to IR problems is the following: 
regarding $\gamma_5$, the delicate diagram is shown
in the first picture of Fig.~(\ref{three-point}) which belong to the pure vertex corrections 
$\Gamma_{\hat{W}^+_\mu \bar{\nu} e}(p_\nu,p_e)$. Luckily, this 
vertex Green function is the simplest one -- compared to 
box and external-leg corrections -- from the IR perspective.  
This is because its physical IR singularities are 
induced only by the on-shell wave-function renormalization.  

Therefore, we can also avoid the physical IR problems by choosing 
an off-shell renormalization procedure. There are two ways of doing 
this:  {\it i)} either one imposes an on-shell renormalization for the 
neutrino and, as a consequence of WTIs, an off-shell 
renormalization prescription of the electron is automatically 
provided (see next section), {\it ii)} or one can 
also choose a MS wave-function renormalization for external 
fermions which is infrared finite and 
compatible with the background gauge invariance. 

Finally, to handle the spurious IR problems generated by means of the 
Taylor subtraction, a modification of the procedure is discussed 
in \cite{amt_2}. Here, the modified breaking terms $\Psi^{(2),W}$, occurring 
in the WTI for the Green function 
$\Gamma_{\hat{W}^+_\mu \bar{\nu} e}(p_\nu,p_e)$, is 
written explicitly. 

\item {\it Kinematic approximations}

The specific kinematic situation allows for some  simplifications:
in $\Gamma_{\hat{W}^+_\mu \bar{\nu} e}(p_\nu,p_e)$
the $W$ is off-shell, while  both the electron and the electron-neutrino 
are on-shell. We can make the 
approximation $p_W^2 = 0$ and neglect 
$m_{muon}/m_W$-terms, because the muon is almost at rest.
All momenta squared are zero, and hence  
%also all three external momenta $p_{W}, p_\nu, p_e$:
\begin{equation}
p_W = p_\nu = p_e = 0 \,.
\label{approx}
\end{equation}
In the following we will derive all symmetry constraints 
for the general kinematic case and then specify to the 
zero-momentum setting~(\ref{approx}).

\item {\it Subdivergences}

Considering only the $O(N_f \alpha^2)$ contributions to 
the two-loop three-point vertex function 
$\Gamma^{(2)}_{\hat{W}^+_\mu \bar{\nu} e}(p_\nu,p_e)$, we have to take 
into account three kinds of one-loop subdivergences: 
the three-gauge boson vertices with one background and two quantum fields 
(they have been largely discussed in \cite{amt_2} within the BFM framework), 
the  quantum gauge boson self-energies (they have been analysed 
in \cite{amt_1,amt_3} without and with the BFM; in particular, in 
\cite{amt_3} the conversion from background field amplitudes to  
quantum ones is completely exploited) and the one-loop 
$\g^{(1)}_{\hat W^+ \bar{\nu} e}(p_\nu,p_e)$ amplitude
together with their corresponding scalar vertices 
where the gauge boson is replaced by the Goldstone boson (notice that 
this amplitude appears also as subdivergence for two-loop three-gauge 
boson vertices and it is extensively discussed in 
\cite{amt_2}). 

Moreover, the renormalization of one-loop amplitudes can 
be quite easily handled within different regularization techniques, 
therefore we can assume, for the time being, that
the one-loop Green functions already satisfy the WTIs (or the STIs) 
and fulfill certain normalization conditions. Nevertheless, to apply 
our procedure, we have to compute the one-loop counterterms which must be inserted 
in one-loop graphs. By using the notation of the introduction, 
these counterterms are given by 
\begin{eqnarray}
  \gg^{(n)}  &=& \gh^{(n)} - \Xi^{(n)} = \g^{(n)} - \left[ T^\delta \g^{(n)} + \
\Xi^{(n)} \right] \nonumber \\
&=&
\g^{(n)}_{\rm bare} - \g^{(n)}_{\rm UV} -  \left[ T^\delta \g^{(n)}_{\rm bare} \
 +
 T^\delta \g^{(n)}_{\rm UV} + \Xi^{(n)} \right] \,. 
  \label{eq:gg_1}
\end{eqnarray}
In the second line we have introduced the bare Green function
$\g_{\rm bare}^{(n)}$ in addition.
This quantity is defined by $\g^{(n)} =
 \g^{(n)}_{\rm bare} - \g^{(n)}_{\rm UV}$ where  $\g^{(n)}_{\rm UV}$
denotes the necessary UV counterterms computed in the specific 
regularization used in the calculation. Of course,
the complete one-loop counterterms, namely $\gg^{(n)} - \g^{(n)}_{\rm bare}$, 
have to be taken into account at the two-loop level.

For instance, in the case of charged gauge-boson self-energies, 
given $\gg^{(1)}_{W^+_\mu  W^-_\nu}(p)$ 
(which satisfy the normalization conditions and the corresponding 
STIs) and given $\g^{(1)}_{W^+_\mu  W^-_\nu}(p)$,  
computed in the same regularization as will be used in the 
two-loop computation of $\Gamma^{(2)}_{\hat{W}^+_\mu \bar{\nu} e}(p_\nu,p_e)$, 
the counterterms are 
\begin{eqnarray}
  \label{cou_1}
  \g^{(1),C.T.}_{W^+_\mu  W^-_\nu}(p) &=& T^2_p 
  \left(\g^{(1)}_{W^+_\mu  W^-_\nu}(p)\right) + \Xi^{(1)}_{W^+_\mu  W^-_\nu}(p)\,,
  \nonumber \\
  \Xi^{(1)}_{W^+_\mu  W^-_\nu}(p) &=& 
    \xi^{(1)}_{ W,1} \, p^2 g_{\mu\nu} +\xi^{(1)}_{ W,2} \, p_\mu p_\nu 
    + \xi^{(1)}_{M_W} g_{\mu\nu}\,, 
\end{eqnarray}
where 
\begin{eqnarray}
  \label{cou_2}
   \xi^{(1)}_{ W,1} &=& {1\over 144} 
   \left( 5\, \partial^2_p 
   \left. \gg^{(1)}_{W^+_\mu  W^-_\mu}(p)\right|_{p=0} - 2\, 
   \partial_{p^\mu} \partial_{p^\nu}           
   \left. \gg^{(1)}_{W^+_\mu  W^-_\nu}(p)\right|_{p=0}\right)\,, \nonumber \\
   \xi^{(1)}_{ W,2} &=& {1\over 72} 
   \left( - \partial^2_p 
   \left. \gg^{(1)}_{W^+_\mu  W^-_\mu}(p)\right|_{p=0} + 
   4 \, \partial_{p^\mu} \partial_{p^\nu}           
   \left. \gg^{(1)}_{W^+_\mu  W^-_\nu}(p)\right|_{p=0}\right)\,,
   \nonumber \\
   \xi^{(1)}_{M_W} &=& {1\over 4}  
   \left. \gg^{(1)}_{W^+_\mu  W^-_\mu}(p)\right|_{p=0}\,. 
\end{eqnarray}
In the same way, all the other possible one-loop 
divergences can be computed once the renormalized 
Green functions $\gg^{(1)}$ are known. 

In some cases, the BFM does not achieve great 
advantages (for instance, in the cases of amplitudes with external 
fermions only) at the practical level and, on the other hand,  
it could be convenient to use the conventional 
gauge fixing. However, the Green functions computed with external background 
fields can be easily related to those with external quantum fields by using 
the extended versions of the BRST symmetry.  

\end{itemize}

\subsection{Two-loop vertex function} 

Working in the framework of the BFM, 
there is only one WTI for the vertex function 
$\Gamma^{(2)}_{\hat{W}^+_\mu \bar{\nu} e}(p_\nu,p_e)$ 
that has to be evaluated at two loops (cf. \cite{amt_1}):
\begin{eqnarray} 
  \label{WTI_1} 
i \left(p_\nu+ p_e \right)_\rho \g^{(2)}_{\hat W^{+}_{\rho} \bar{\nu} 
    e}(p_\nu,p_e)      
  +  i\, M_W  \g^{(2)}_{\hat G^{+} \bar{\nu} e}(p_\nu, p_e)  
\hphantom{xxxxxxxxxx} 
\nonumber   \\ \mbox{} 
 + { i e \over s_W\sqrt{2}} \left[ 
    \g^{(2)}_{\bar{\nu} \nu}(- p_\nu)  
   P_L 
    -  P_R 
    \g^{(2)}_{\bar{e} e}(p_e)   \right] &=&   
  \Delta^{(2),W}_{\lambda_+ \bar{\nu} e}(p_\nu,p_e) 
\,.\nonumber \\
\end{eqnarray} 
Here $\Delta^{(2),W}_{\lambda_+ \bar{\nu} e}(p_\nu,p_e)$ 
is a polynomial of the external momenta 
$p_\nu$ and $p_e$ of maximum degree 1.
%No further contraint, for example on the amplitude 
%$\Gamma_{\hat{G}^+,\bar{p},p'}(p,p')$  comes into the play.
We define the weak mixing angle through 
the on-shell relation $c_W=M_W/M_Z$ 
as we want to maintain the form of the WTIs to be the 
same to all orders. $P_{L / R}=(1\mp\gamma_5)/2$ are the 
chiral projectors.

The breaking terms in~(\ref{WTI_1}) are generated by a non-invariant 
regularization procedure, for instance by using the 
`t Hooft-Veltman definition 
of $\gamma_5$ \cite{hoo}. To remove them,  according to the 
procedure described in \cite{amt_1,amt_2}, 
we apply the Taylor operator $(1-T^1_{p_\nu,p_e})$, obtaining 
\begin{eqnarray} 
  \label{WTI_2} 
 &&\hspace{-.5cm} i \left(p_\nu + p_e \right)_\rho
\left[\left(1-T^0_{p_\nu,p_e}\right) 
\g^{(2)}_{\hat W^{+}_{\rho} \bar{p_\nu} 
    p_e}(p_\nu , p_e)\right]      
  +  i\, M_W \left[\left(1-T^0_{p_\nu,p_e}\right) 
\g^{(2)}_{\hat G^{+} \bar{\nu} e}(p_\nu, p_e) \right]  
  %\hphantom{xxxx} 
  \nonumber \\ 
 && + { i e \over s_W\sqrt{2}} \left\{ 
   \left[\left(1-T^1_{p_\nu}\right) \g^{(2)}_{\bar{\nu} \nu}(- p_\nu)\right]  
   P_L -  P_R
   \left[\left(1-T^1_{p_e}\right) \g^{(2)}_{\bar{e} e}(p_e)\right]  
 \right\}  
  =   
  \Psi^{(2),W}_{\lambda_+  \bar{\nu} e}(p_\nu,p_e) 
  \,. \nonumber \\
\end{eqnarray} 
where  
\begin{eqnarray} 
  \Psi^{(2),W}_{\lambda_+  \bar{\nu} e}(p_\nu,p_e) = 
  i\, M_W  \left( p^\rho_\nu \partial_{p^\rho_\nu} + {p}^\rho_e 
    \partial_{p^\rho_e} \right)   
  \g^{(2)}_{\hat G^{+} \bar{\nu} p_e}(p_\nu,p_e) \Bigg|_{p_\nu=p_e=0} \,,
  \label{WTI_3} 
\end{eqnarray} 
are finite and are generated by means of the over-subtraction. 

Notice that the computation of $\Psi^{(2),W}_{\lambda_+ \bar{\nu} e}$ 
can be also performed without encountering any UV divergences. For 
that purpose, it is sufficient to implement the subtraction 
of UV subdivergences directly in such a way that all the diagrams are 
always finite. Technically, we suggest the use of Zimmermann's 
subtraction formula \cite{zimm}. Since all the integrals involved 
can be performed analytically, $\Psi^{(2),W}_{\lambda_+ \bar{\nu} e}$ 
can be evaluated exactly. In this way, we use
the Dirac algebra in $4$ dimensions without any ambiguities. 

Notice that the zero momentum subtraction in Eq.~(\ref{WTI_2}) 
removes exactly the contribution that we would like 
to evaluate to compute the muon decay amplitude 
in the approximation~(\ref{approx}). However, this contribution 
can be computed as a counter\-term. Notice in fact that, 
the local part of the muon decay amplitude is totally fixed when  
the normalization conditions for fermion two-point functions and the 
WTIs are used. 

By using the parametrization
\begin{eqnarray}\label{par_1} 
  \Xi^{(2)}_{\bar\psi \psi}(p) &=&  
  \xi^{(2)}_{2,\psi} \left(\not\!p - m_\psi \right) + 
  \xi^{(2)}_{\psi} m_\psi \,,~~~~~\psi=\nu,e \nonumber \\
  \Xi^{(2)}_{\hat W^{+}_{\mu} \bar{\nu} e}(p_\nu,p_e)  
  &=&  \xi^{(2)}_{L, \hat W^{+} \bar{\nu} e} \, \gamma^\mu   P_L\,   
       + \xi^{(2)}_{R, \hat W^{+} \bar{\nu} e} \, \gamma^\mu   P_R \,, 
\end{eqnarray} 
for two- and three-point functions, and decomposing the 
breaking terms into scalar functions
\begin{eqnarray}\label{par_1} 
  \Psi^{(2),W}_{\lambda_+  \bar{\nu} e}(p_\nu,p_e)= i\left( 
  \psi^{(2)}_1\not\!p_\nu  P_L  + 
  \psi^{(2)}_2 \not\!p_\nu  P_R + 
  \psi^{(2)}_3 \not\!p_e P_L  + 
  \psi^{(2)}_4 \not\!p_e  P_R \right)   
  \,,
\end{eqnarray} 
we have the solution 
\begin{eqnarray}\label{xi_sol}
\xi^{(2)}_{R, \hat W^{+} \bar{\nu} e} &=& 
\psi^{(2)}_2 =   \psi^{(2)}_4\,, \nonumber \\ 
 \xi^{(2)}_{L, \hat W^{+} \bar{\nu} e} &=& 
{e \over s_W \sqrt{2}} \, \xi^{(2)}_{2,\nu} + \psi^{(2)}_1\,, \nonumber \\
\xi^{(2)}_{2,e} &=& \xi^{(2)}_{2,\nu} + {s_W \sqrt{2} \over e} 
\left(\psi^{(2)}_1 - \psi^{(2)}_3\right)\,,
 \end{eqnarray}
for the coefficients. Notice that the equality $\psi^{(2)}_2 =   \psi^{(2)}_4$ follows from 
the consistency conditions (see, for example, the discussion in \cite{amt_1}) and it 
provides a check of the computation of the breaking terms. In addition, from 
Eqs.~(\ref{xi_sol}), it emerges that $\xi^{(2)}_{2,e} = \xi^{(2)}_{2,\nu}$ 
in the case of invariant regularization techniques, namely when $\psi^{(2)}_i = 0,~~\forall i$. 
This means that we are not allowed to impose any arbitrary normalization conditions for fermion residues. If the 
neutrino is renormalized in such a way that its residues is equal to 1, the electron 
residue will be clearly different from 1. In this way, we have a partial on-shell scheme, 
we maintain the background gauge symmetry and we can avoid the physical IR divergences 
for the vertex amplitude. 
 
The final result, namely diagram computation plus counterterms, can be 
written in the following way
\begin{eqnarray} 
  \gg^{(2)}_{\hat W^{+}_{\mu} \bar{\nu} e}(p_\nu,p_e)   
  &=&  
  \g^{(2)}_{\hat W^{+}_{\mu} \bar{\nu} e}(p_\nu,p_e) 
  \nonumber\\&&\mbox{} 
  - \left[ T^0_{p_\nu,p_e} \g^{(2)}_{\hat W^{+}_{\mu} \bar{\nu} e}(p_\nu,p_e)   
  + \xi^{(2)}_{L, \hat W^{+}\bar{\nu} e} \gamma^\mu P_L 
  + \xi^{(2)}_{R, \hat W^{+}\bar{\nu} e} \gamma^\mu P_R \right]  
 % \nonumber\\&&\mbox{} 
 % -  {e  \over 2 \sqrt{2} s_W} \left(\xi^{(2)}_{2,\nu} + \xi^{(2)}_{2,e} 
  %\right) \gamma^\mu P_L  
  \,.   
  \label{counter3} 
\end{eqnarray} 
The parameters $\xi^{(2)}_{\nu}$ and $\xi^{(2)}_{e}$ 
are used to impose mass renormalization conditions on the fermion self-energies.  

From this we learn that the symmetric amplitude 
$\gg^{(2)}_{\hat W^{+}_{\mu} \bar{\nu} e}$ 
at zero momentum is just given by the universal counterterm:
\begin{eqnarray} 
  \gg^{(2)}_{\hat W^{+}_{\mu} \bar{\nu} e}(p_\nu=0,p_e=0)   
  &=&  
  - \xi^{(2)}_{L, \hat W^{+}\bar{\nu} e} \gamma^\mu P_L 
  - \xi^{(2)}_{R, \hat W^{+}\bar{\nu} e} \gamma^\mu P_R   
  %\nonumber\\&&\mbox{} 
  %-  {e  \over 2 \sqrt{2} s_W} \left(\xi^{(2)}_{2,\nu} 
  %  + \xi^{(2)}_{2,e} 
  %\right) \gamma^\mu P_L
\,.   
  \label{counter4} 
\end{eqnarray} 
The proposed procedure allows for an efficient computation of the 
amplitude $\gg^{(2)}_{\hat W^{+}_{\mu} \bar{\nu} e}$ at two-loop 
order avoiding the $\gamma_5$ problem. In the literature, different techniques with different 
prescription of $\gamma_5$ have been used to evaluate the 
Feynman diagrams of $\gg^{(2)}_{\hat W^{+}_{\mu} \bar{\nu} e}$, however we believe that 
a rigorous check of these result is desirable. 

%In the computation of the final amplitude $\xi^{(2)}_{2,\nu}$ 
%will drop out, since it takes into account only the wave function 
%renormalization of the neutrino.
%
%The electron mass can be safely neglected in the calculation.
%However, in this approximation, the Taylor subtraction at 
%zero-momentum might lead to spurious IR divergences. 
%Following the general procedure presented in Section 6 
%of~\cite{amt_2}, we should modify slightly the general strategy 
%in such a way that the breaking terms are IR-finite.
%In our example only the highest order of the Taylor derivative 
%leads to IR divergences. Therefore, we can modify our procedure by
%replacing~(\ref{WTI_3}) by
%\begin{eqnarray} 
%  \Psi'^{(2)}_{\lambda_+  \bar{\nu} e}(p_\nu,p_e)   &=&  
%  i\, M_W   
%\left( T^1_{p_\nu p_e} - T^0_{p_\nu p_e} \right)   
%    \g^{(2)}_{\hat G^{+} \bar{\nu} e}(p_\nu,p_e)  
%  \nonumber \\&&\mbox{} 
%  - { i e \over s_W\sqrt{2}}   
%    (T^1_{p_\nu}-T^0_{p_\nu}) 
%   \g^{(2)}_{\bar{\nu} \nu}(-p_\nu)P_L +  P_R {i e \over s_W\sqrt{2}}   
%    (T^1_{p_e}-T^0_{p_e}) 
%   \g^{(2)}_{\bar{e} e}(p_e)
%  \,.\nonumber \\ 
%  \label{WTI4IR} 
%\end{eqnarray} 
%Notice that  the advantages due to the zero-momentum subtractions 
%have only slightly been reduced. The computation of the breaking terms  
%still relies on Green functions expanded around zero external momenta. 
%We also stress that the proposed rearrangement solves the spurious 
%IR problem due to Taylor subtractions in general. 

%%%%%%%%%%%%%%%%%%%%%%%%%%%%%%%%%%%%%%%%%%%%%%%%%%%%%%%%%
 
\section{Massive U(1) BRST symmetry within the SM}
\label{IR}

It is a well-known problem that in the computation of Green's function
there are IR divergences due to the vanishing photon mass.  In this
brief section, we will describe how to perform a regularization of the
photonic IR divergences in a way that is consistent with the BRST
symmetry of the SM and, thus, preserves the unitarity of the model
(see also \cite{g}). The choice of such a regulator is motivated essentially by 
the fact that it regulates both the physical and spurious IR divergences.

We will refer to the Stueckelberg method (see
\cite{stuck,DHN} and references therein).  This method gives rise to
unsolvable renormalization problems in the case of Yang-Mills fields,
which only can only be resolved through the Higgs mechanism. However, in
the abelian case it provides a manifestly BRST (and the background
gauge) invariant model of massive QED.

For pedagogical purposes, we present a short digression regarding 
the Stueckelberg formalism in QED. The Lagrangian is given by 
\begin{eqnarray}\label{QED_stu}
{\cal L} = -\frac{1}{4 g^2}F_{\mu\nu}{}^2  +
\frac{m^2}{2g^2}(A_\mu -\frac{1}{m}\partial_\mu \varphi)^2 
+ s( \bar c {\cal F} ) + {\cal L}_{\rm matter}\,, \nonumber \\
{\cal L}_{\rm matter} = \bar{\Psi}(i\not\!\partial-M)\Psi + 
\bar{\Psi}\gamma^\mu(A_\mu-\frac{1}{m}\partial_\mu\varphi)\Psi\,,
\end{eqnarray}
where $F_{\mu\nu} = \partial_\mu {A}_\nu - \partial_\nu {A}_\mu$.

The apparent non-renormalizable derivative coupling  
$\bar{\Psi}\gamma^\mu\partial_\mu\varphi\Psi$ can be absorbed 
by a field redefinition
\begin{equation}\label{red}
\psi=e^{\frac{i}{m}\varphi }\Psi, \qquad
{\cal L}_{matter}= \bar{\psi}(\gamma^\mu(i\partial_\mu + A_\mu) -M)\psi.
\end{equation}
Notice that in the non-abelian case a redefinition of fields of type (\ref{red})
will generate new non-eliminable non-renormalizable terms \cite{DHN}.  

The BRST transformations of the fundamental fields are given by 
\begin{eqnarray}
\nonumber
s \hat{A}_\mu&=0\,,  \quad \, \, \hat{A}_\mu &= A_\mu -\frac{1}{m}\partial_\mu\varphi \,,\\
\label{abelian}
s \varphi &=m\, c \,,\quad s c &=0\,,\\
\nonumber
s \bar c &= b \,,\quad s \, b&=0 \,.
\end{eqnarray}
Thus, the BRST multiplets consist of two trivial pairs 
$(\varphi, c)$, $(\bar c, b)$ and one
singlet $\hat{A}_\mu = A_\mu-\frac{1}{m}\partial_\mu\varphi$. This means 
that the physical spectrum will be independent of $(\varphi, c)$ and 
$(\bar c, b)$. 

In the gauge-fixing-ghost term $ s( \bar c {\cal F} )$, 
${\cal F}$ is a real bosonic function of all the fields and their derivatives. 
For example, the `t Hooft-Feynman gauge fixing 
\begin{equation}
{\cal F} = \frac{1}{g^2}(\frac{1}{2}b - \partial^\mu A_\mu - m \varphi)\,,
\end{equation}
leads directly to noninteracting ghosts $\bar c$ and $c$. Moreover,  
the $\hat{A}-\varphi$ sector does not contain higher derivatives or dipoles.
The field strengths $F_{\mu\nu}$ calculated from $A$ and from 
$\hat{A}$ coincide. The field $b$ is auxiliary with the algebraic equation of motion
\mbox{$b = \partial^\mu {A}_\mu + m \varphi $} 
Thus, the ${\cal L}$ defines a massive abelian gauge field coupled to matter fields.

To extend the Stueckelberg formalism to the SM quantized in the 
background gauge, the scalar field $\varphi$ transforms under 
the BRST symmetry\footnote{Notation, definitions,
and quantum numbers can be found in \cite{amt_1}.}
and under the background gauge transformations in the following way
\begin{equation}
  \label{eq:tras}
  s \, \varphi = \mu \, c \,, ~~~~~~  \delta_{(\lambda)} 
  \varphi = \mu \, \lambda \,,  
\end{equation}
where $\lambda$ is
the infinitesimal parameter of the background gauge transformations. 
$\mu$ is the IR regulator and $c$ is the $U(1)$ ghost. The
latter can be written in terms of the combination $ c= c_W c_A + s_W
c_Z $, where $c_A$ and $c_Z$ are the photon ghost and the ghost
associated with the $Z$ boson, respectively.  It follows that a term
like
\begin{equation}\label{sch2} 
\Gamma^{\rm Stu} =  
\int d^{4}x 
\left( {1\over 2} \partial_{\mu}\varphi  \partial^{\mu} \varphi  - 
\mu \partial_{\mu} \varphi B^{\mu} +  {\mu^{2}\over 2} B^{\mu} B_{\mu} \right) \,,
\end{equation}
is BRST and background gauge invariant for all the values of the
parameter $\mu$ and the last term provides a mass term for the $B_\mu$
fields.  Other $\varphi$-dependent invariant terms can be constructed,
however it is easy to show that all of them, but (\ref{sch2}), can be
reabsorbed by a simple redefinition of the field $B_\mu$ (cf.
\cite{DHN}). In addition, in order to deal with diagonal two-point
functions the `t Hooft gauge fixing
\begin{eqnarray}
  \label{sch3}
  \Gamma^{\rm Stu, g.f.} 
&=&  s \, \int d^{4}x \, \bar c \, \Big( \partial^\mu B_\mu - \rho \mu \, \varphi  + 
\xi_0 b \Big)  \nonumber \\
&=& \int d^4x \left[ b  \left( \partial^\mu B_\mu - \rho \mu \, \varphi  + \xi_0 b \right) - 
\bar c \, \partial^2  c + \rho \mu^2 \bar c \, c 
\right]\,,
\end{eqnarray}
is used. Here $\xi_0$ is the conventional gauge fixing parameter of the $U(1)$ sector and 
$\rho$ is the `t Hooft parameter. With this gauge fixing, it is easy to see that the gauge 
field $B_\mu$, the scalar field $\varphi$ 
and the ghosts $\bar c, c$ (with masses $\mu^2, \rho^2 \mu^2/ \xi_0$ and 
$\rho \mu^2$) form a quartet which ensures the unitarity of the model. Notice that the 
BRST variation of $\varphi$ says 
that this field corresponds to a would-be-Goldstone boson, and the spontaneous 
symmetry breaking mechanism -- in the abelian case -- can be implemented without 
the Higgs counterpart. 

In the SM framework, the field $B_\mu$ does not coincide with 
the physical photon field, but the mixing with the third component of 
$SU(2)$ gauge boson triplet $W^3_\mu$ has to be considered. 
We have to cancel this term by modifying  the gauge fixing 
function ${\cal F}_{B}$ (cf. \cite{amt_1}, Eq.~(A.3))  for the abelian field 
in the following way
\begin{equation}\label{gauf}
{\cal F}_{B}= \partial^\mu B_\mu + \rho_0 (\hat \Phi + v )^i t^0_{ij} 
\left( \Phi + v\right)^j + \frac{\xi_0}{2} \, b
 \longrightarrow {\cal F}_{B} - \rho  \mu \varphi \,,
\end{equation} 
where $\rho$ is the t'Hooft parameter for the $\varphi$ field. 
By eliminating the Lagrange multiplier $b$, we have the 
gauge fixing terms
\begin{eqnarray}
  \label{gauf_1}
{\cal L}^{\rm g.f.} &=&  -\frac{1}{2 \xi_0} \left[ \partial^\mu B_\mu   - \rho  \mu \varphi  
+ \rho_0 ( \hat \Phi + v )^i t^0_{ij}  \left( \Phi + v\right)^j \right]^2
 \nonumber \\
&=&
-\frac{1}{2 \xi_0} (\partial^\mu B_\mu )^2 - { \rho^2 \mu^2 \over 2 \xi_0 }  \varphi^2 - 
{\rho_0 g'^2 v^2 \over 2 \xi_0} G^2 \nonumber 
+ {\rho \rho_0 \mu v g' \over 2 \xi_0} \varphi \, G \nonumber \\
& & +  
{\rho \mu  \over \xi_0 } \partial^\mu B_\mu \, \varphi - { \rho_0 v g' \over \xi_0 } 
\partial^\mu B_\mu G \nonumber \\
&& 
- {\rho v g' \over 2 \xi_0}\left( \partial^\mu B_\mu -  \mu \rho \varphi \right)  
\left( H \, \hat G - \hat H \, G \right)
- {\rho^2 g'^2 v^2 \over 2 \xi_0} \left( H \, \hat G - \hat H \, G \right)^2 
\,,  
\end{eqnarray}
where $g'$ is the $U(1)$ gauge coupling, $v$ is the vacuum expectation
value, $G$ and $H$ are the Goldstone boson and the Higgs field,
respectively, while $\hat G$ and $\hat H$ are their background partners.  
The first line contains the contribution to the quadratic
part of the action, this shows that also the masses of the Goldstone
boson $G$ are modified by the introduction of the Stueckelberg field
$\varphi$. The mixed terms $ \varphi \partial^\mu B_\mu$ and 
$G \partial^\mu B_\mu$ are cancelled (in the restricted `t Hooft gauge) by the
mixing terms coming from the covariant derivatives of the kinetic
terms ({i.e.}\, from Eq.~(\ref{sch2})). Finally the last terms
describe the interactions between $\varphi$ and the other fields. As
can be noticed all the interaction terms depend on the background
fields.  Therefore,  the Stueckelberg $\varphi$ field can be generated only if
the fields $\hat G$ or $\hat H$ appear as external vertices of the
amplitude or due to the mixing with the neutral Goldstone boson. 
This is the only difference between the Stueckelberg formalism and its
application to the SM quantized in the background gauge.

\Acknowledgements
We thank P.~Gambino, W.~Hollik, G.~Weiglein, A.~Ferroglia and M.~Pernici 
for useful comments and discussions. We also  warmly thank our collaborator Matthias Steinhauser. 
The work of P.A.G. is supported by NSF grants no. PHY-9722083 and PHY-0070787.

%%%%%%%%%%%%%%%%%%%%%%%%%%%%%%%%%%%%%%%%%%%%%%%%%%

%  
% references  %%%%%%%%%%%%%%%%%%%%%%%%%%%%%%%%%%%%%%%%
%  


\begin{thebibliography}{99}

 \bibitem{pige}
  O. Piguet and S.P. Sorella, {\it Algebraic Renormalization}   
  Lecture Notes in Physics Monographs, Springer-Verlag  
  Berlin Heidelberg, 1995 and references therein.

\bibitem{amt_1} 
  P.A.~Grassi, T.~Hurth, and M.~Steinhauser,  [hep-ph/9907426] to appear in Annals of Physics (NY).

\bibitem{amt_3} 
  P.A.~Grassi, T.~Hurth, and M.~Steinhauser,  Rep. No. CERN-TH/2001-002, NYU-TH/00/09/9. 

\bibitem{amt_2} 
  P.A.~Grassi, T.~Hurth, and M.~Steinhauser,  [hep-ph/0011067] to appear in JHEP.

\bibitem{bkg}   
  G.~`t Hooft,  Nucl. Phys. {\bf B33}, {436} {(1971)};
  H.~Kluberg-Stern and J.~Zuber, Phys. Rev. D {\bf 12}, {467} {(1975)}; 
  Phys. Rev. D {\bf 12}, {482} {(1975)};
  L.F.~Abbott,  Nucl. Phys. {\bf B185}, {189} {(1981)};
  S.~Ichinose and M.~Omote, Nucl. Phys. {\bf B203}, {221} {(1982)};
  D.M.~Capper and A.~MacLean, Nucl. Phys. {\bf B203}, {413} {(1982)};   
  D.G.~Boulware, Phys. Rev. D {\bf 12}, {389} {(1981)}.

\bibitem{msbkg} 
  A.~Denner, G.~Weiglein, and S.~Dittmaier, Phys. Lett. {\bf B333}, {420} {(1994)}; 
  Nucl. Phys. {\bf B440}, {95} {(1995)}.

\bibitem{grassi}  
  P.A.~Grassi, Nucl. Phys. {\bf B462}, {524} {(1996)}; 
  Nucl. Phys. {\bf B537}, {527}{(1999)}; Nucl. Phys. {\bf B560}, {499} {(1999)}.

\bibitem{sirlin}
A.~Sirlin,
%``Radiative Corrections In The SU(2)-L X U(1) Theory: A Simple Renormalization Framework,''
Phys.\ Rev.\ D {\bf 22},  971 (1980). 
A.~Sirlin,
%``On The O (Alpha**2) Corrections To Tau (Mu), M (W), M (Z) In The SU(2)-L X U(1) Theory,''
Phys.\ Rev.\ D {\bf 29},  89 (1984).

\bibitem{stuart1}
T.~van Ritbergen and R.~G.~Stuart,
%``Complete 2-loop quantum electrodynamic contributions to the muon  lifetime 
%in the Fermi model,''
Phys.\ Rev.\ Lett.\ {\bf 82}, 488 (1999) [hep-ph/9808283].
T.~van Ritbergen and R.~G.~Stuart,
%``On the precise determination of the Fermi coupling constant from the  muon lifetime,''
Nucl.\ Phys.\ {\bf B564}, 343 (2000) [hep-ph/9904240]. 

\bibitem{mat}
M.~Steinhauser and T.~Seidensticker,
%``Second order corrections to the muon lifetime and the semileptonic B  decay,''
Phys.\ Lett.\ {\bf B467}, 271 (1999)
[hep-ph/9909436].

\bibitem{ferroglia}
A.~Ferroglia, G.~Ossola, and A.~Sirlin,
%``Considerations concerning the radiative corrections to muon decay in  the 
%Fermi and standard theories,''
Nucl.\ Phys.\ {\bf B560}, 23 (1999) [hep-ph/9905442].

\bibitem{stuart2}
P.~Malde and R.~G.~Stuart,
%``Complete O(N(f) alpha**2) weak contributions to the muon lifetime,''
Nucl.\ Phys.\ {\bf B552}, 41 (1999) [hep-ph/9903403].

\bibitem{weig1}
A.~Freitas, W.~Hollik, W.~Walter, and G.~Weiglein,
%``Complete fermionic two-loop results for the M(W) - M(Z)  interdependence,''
Phys.\ Lett.\ {\bf B495}, 338 (2000) [hep-ph/0007091].

\bibitem{weig2}
A.~Freitas, S.~Heinemeyer, W.~Hollik, W.~Walter, and G.~Weiglein,
%``Calculation of fermionic two-loop contributions to muon decay,''
Nucl.\ Phys.\ Proc.\ Suppl.\ {\bf 89}, 82 (2000) [hep-ph/0007129].

\bibitem{hoo}
  G. `t~Hooft and M. Veltman, Nucl. Phys. {\bf B44}, {189} {(1972)}; \, 
  P.~Breitenlohner and D.~Maison, Comm. Math. Phys. {\bf 52}, {11, 39, 55} {(1977)}. 

\bibitem{zimm}
  W. Zimmermann, 
  {\it Local Operator Products and Renormalization in   
    Quantum Field Theory} \,   
  in 1970 Brandeis University Summer Institute Lectures,   
  Cambridge, Mass. M.I.T. Press.

\bibitem{brs}  
  C. Becchi, A. Rouet, and R. Stora, Comm. Math. Phys. {\bf 42}, {127}{(1975)}; 
Ann. of Phys. (NY) {\bf 98}, {287}{(1976)};
  I.V. Tyutin, Lebedev Institute preprint N39 (1975); see also
  C. Becchi, {\it Lectures on Renormalization of   
    Gauge Theories} in Relativity, group and topology II, Les Houches 1983.

\bibitem{g} P.A. Grassi, Ph. D. Thesis, University of Milano, Italy, 1997. 

\bibitem{stuck} E.C.G. Stueckelberg, Helv. Phys. Acta {\bf 30}, 209  (1957). 

\bibitem{DHN} 
 N.~Dragon, T.~Hurth, and P.~van Nieuwenhuizen,
{Nucl.\ Phys.\ Proc.\ Suppl.} {\bf B56}, 318  (1997) [hep-th/9703017];
 T.~Hurth, { Helv.~Phys.~Acta} {\bf 70},  406 (1997).

\end{thebibliography}
\end{document}